\documentclass[10pt,letterpaper]{article}
\usepackage{amsmath}
\usepackage{amssymb}
\usepackage{cite}

\usepackage{opex3}
\usepackage{txfonts}
\usepackage{amssymb}
\usepackage{tipa}
\usepackage{graphicx}
\usepackage{txfonts}
\usepackage{amssymb}
\usepackage{extarrows}
\begin{document}

\title{Phase sensitive photonic flash}
\author{X. Y. Cui,$^{1}$ J. H. Wu,$^{1}$, and Z. H. Wang,$^{1,2,*}$ }
\address{$^1$ Center for Quantum Sciences and School of Physics and Center for Advanced Optoelectronic Functional Materials Research and Key Laboratory for UV Light-Emitting Materials and Technology of Ministry of Education, Northeast Normal University, Changchun 130024, China\\
$^2$ Beijing Computational Science Research Center, Beijing 100094, China }
\email{$^{*}$wangzh761@nenu.edu.cn}

\begin{abstract}
We theoretically propose a photonic flash based on a linearly coupled cavity system. Via driving the two side cavities by external fields, it forms a cyclic energy-level diagram and therefore  the phase difference between the driving fields acts as a controller of the steady state due to the quantum interference effect.  In the optical trimer structure, we show that the perfect photonic flash can be realized in the situation of resonant driving. The perfect photonic flash scheme is furthermore generalized to multiple coupled cavity system, where the cavities with odd and even number turn bright and dark alternatively.  {Our proposal may be applied in the designing of quantum neon and realizing a controllable photonic localization.}
\end{abstract}



\section{Introduction}
Coherent photonic control and designing photonic devices are of potential application in quantum information processing, and the coupled cavity array system provides an ideal platform for achieving such tasks.

In the coupled cavity array system, which can be realized by the photonic crystal~\cite{MB}, toroid micocavity~\cite{DK}  as well as superconductive transmission line~\cite{AW,AB}, the coherent photonic control has invoked a lot of attentions. In the ideal situation, where the decay of the cavities is neglected, kinds of single-photon devices have been theoretically proposed within the scattering frame, such as quantum transistor~\cite{ZL1}, quantum router~\cite{ZL2} and quantum frequency converter~\cite{wang,zheng,yan}. On the other hand, in the coupled cavity array with loss, it has been shown that the non-equilibrium quantum phase transition, which is related to
the steady state, may occur in some nonlinear system when the driving and dissipation are both present~\cite{FN,AL,JJ1,JJ,JR,JJM}. Therefore, it supplies a photonic platform to simulate many body phenomenon in condensed matter physics~\cite{CN,SS}. Naturally, the steady state in the coupled cavity array with only linear interaction deserves more exploration.

In this paper, by investigating the photonic distribution of steady state in a linearly coupled cavity array system, we propose a scheme to realize photonic flash. We here use the phrase ``photonic flash'' to refer that  when some of cavities in the coupled cavity array are completely dark, the other ones are completely bright, and the bright and dark cavities can be exchanged by tunning the system parameters appropriately. In our scheme, the two side cavities are driven by a pair of external fields, we will show that the phase difference between the driving fields acts as an ideal controller for the steady state, which stems from the quantum interference effect among the multiple transition channels. As a simple example, we sketch the realization of dark cavity in an optical trimer, and show that we can construct a perfect photonic flash under the situations (i) the central cavity of the trimer is lossless and (ii) the two side cavities are driven resonantly. Generalizing the scheme to an array of multiple coupled cavities, we show that the cavities with odd and even number will be bright and dark alternatively by adjusting the phase difference. Therefore, we design a scalable photonic flash based on linear coupled cavity array system. Instead of controlling each cavity individually, we here only drive the two side cavities, regardless of the total number of the cavities, and therefore is beneficial to avoiding the difficulties of addressing. { As the potential application in the designing of photonic device and quantum simulation  based on our proposal, we also give a simple remark on the realizing of quantum neon and controllable photonic localization.}

The rest of the paper is organized as follows. In Sec.~\ref{model}, we present the model of an optical trimer and the steady state values. In Sec.~\ref{flashtrimer}, we discuss the condition to realize perfect photonic flash in the trimer structure and generalize it to multi-coupled-cavity system in Sec.~\ref{generalize}.  At last, we discuss the effect of quantum noise and give a brief remark and conclusion in Sec.~\ref{remark}.

\begin{figure}[tbp]
\centering
\includegraphics[width=0.8\columnwidth]{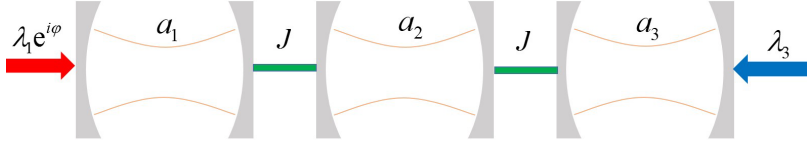}
\caption{The scheme of an optical trimer for photonic flash, where the two side cavities are driven by classical fields.}
\label{trimmer}
\end{figure}

\section{Model of optical trimer}
\label{model}
The optical trimer system in our consideration is sketched in Fig.~\ref{trimmer}, in which
the cavity $1,3$ are coupled to the cavity $2$  simultaneously and are driven by classical fields with the same frequency $\omega_d$.  In the rotating frame with respect to the frequency of the driving fields, the Hamiltonian can be written as (here and after $\hbar=1$)
\begin{eqnarray}
H&=&\sum_{i=1}^{3} \Delta_{i}a_{i}^{\dagger}a_{i}+J(a_{1}^{\dagger}a_{2}+a_{2}^{\dagger}a_{1}
+a_{2}^{\dagger}a_{3}+a_{3}^{\dagger}a_{2})\nonumber\\&&
+\lambda_{1}(a_{1}e^{i\varphi}+a_{1}^{\dagger}e^{-i\varphi})
+\lambda_{3}(a_{3}+a_{3}^{\dagger}),
\end{eqnarray}
where $\Delta_i=\omega_i-\omega_d\,(i=1,2,3)$ is the detuning between the $i$th cavity and the driving fields. Here, $\omega_i$ is the resonant frequency of cavity
$i$, which is described by the annihilation operator $a_i$. $J$ is nearest inter-cavity coupling strength. $\lambda_1$ and $\lambda_3$ are the driving strengths to cavity $1$ and $3$, respectively. $\varphi$ is the phase difference between the two driving fields, which will play an important role in the controlling of photons. Without loss of generality, we assume that all the parameters are real.

Assuming the decay rate of the $i$th cavity is $\gamma_i(>0)$, the dynamics of the system is determined by the Heisenberg-Langevin equation (neglecting
the fluctuations)
\begin{equation}
\dot{\bf A}=M{\bf A}+\bf{B},
\end{equation}
where ${\bf A}=(a_1,a_2,a_3)^T, {\bf B}=(-i\lambda_{1}e^{-i\varphi},0,-i\lambda_{3})^T$, and the matrix $M$ is
\begin{equation}
M=\left(\begin{array}{ccc}
-i\Delta_{1}-\gamma_{1}/2&-iJ&0\\
-iJ&-i\Delta_{2}-\gamma_{2}/2&-iJ\\
0&-iJ&-i\Delta_{3}-\gamma_{3}/2
\end{array}\right).
\label{MM}
\end{equation}
Then, the steady-state values of the system are immediately given by
\begin{subequations}
\begin{eqnarray}
\langle a_{1}\rangle&=&\frac{iJ^{2}[
\lambda_3-\lambda_1e^{-i\varphi}(1+K_3)]}
{\rm{det}(M)}, \\
\langle a_{2}\rangle&=&\frac{-J[\lambda_{1}e^{-i\varphi}(i\Delta_{3}
+\gamma_{3}/2)+\lambda_{3}(i\Delta_{1}+\gamma_{1}/2)]}{\rm{det}(M)},\nonumber\\ \\
\langle a_{3}\rangle&=&\frac{iJ^{2}[
\lambda_1e^{-i\varphi}-\lambda_3(1+K_1)]}
{\rm{det}(M)}.
\end{eqnarray}
\label{average}
\end{subequations}
where $K_m:=(i\Delta_{2}+\gamma_{2}/2)(i\Delta_{m}+\gamma_{m}/2)/J^2,\,m=1,3$. Therefore, the steady state property is sensitively dependent on the phase difference $\varphi$ between the two driving fields. This phase dependence mechanism is similar to those in optical molecule (dimer) system~\cite{wang2}, superconducting artificial~\cite{liu,Jia}, chiral molecule~\cite{MS, MS1,MS2,YL}, cavity-QED system~\cite{JT,YLiu} and cavity
optomechanical system~\cite{Jia1,Xu}, where the coupling and driving contribute a cyclic energy-level structure.

\section{Photonic flash in the trimer}
\label{flashtrimer}
As discussed above, the phase difference is a potential controller for the photonic state in an optical trimer, which supplies us a convenient way to design the coherent optical device, such as photonic flash, utilizing the interference effect. Here, we use the phrase ``photonic flash'' to mean that, by adjusting only the phase difference between the driving fields, we will make one or few of the cavities reaches their vacuum steady state, while the intensity in other cavities reach their maximum values. In other words, we will realize the bright and dark cavities simultaneously, and they can be controllably swapped.

To be explicit, we consider a simple scheme in which $\Delta_{1}=\Delta_{2}=\Delta_{3}=\Delta,\lambda_{1}=\lambda_{2}=\lambda_{3}
=\lambda,\gamma_{1}=\gamma_{3}=\gamma,\gamma_{2}=\gamma'$. According to the formula in Eqs.~(\ref{average}), the condition for $\langle a_1\rangle=0$ is
\begin{equation}
\cos\varphi=\frac{4J^{2}+\gamma\gamma'-4\Delta^{2}}{4J^{2}},\,
\sin\varphi=\frac{\Delta\gamma+\Delta\gamma'}{2J^{2}},
\end{equation}
and the condition for $\langle a_3\rangle=0$ is
\begin{equation}
\cos\varphi=\frac{4J^{2}+\gamma\gamma'-4\Delta^{2}}{4J^{2}},\,
\sin\varphi=-\frac{\Delta\gamma+\Delta\gamma'}{2J^{2}}.
\end{equation}
Obviously, the above two equations imply
\begin{equation}
\left(4J^{2}+\gamma\gamma'-4\Delta^{2}\right)^{2}
+4\left(\Delta\gamma+\Delta\gamma'\right)^{2}=16J^{4}.
\label{condition}
\end{equation}
Furthermore, in the case of $\varphi=\pi$, we will obtain that
$\langle a_2\rangle=0$.

\begin{figure}[tbp]
\centering
\includegraphics[width=0.8\columnwidth]{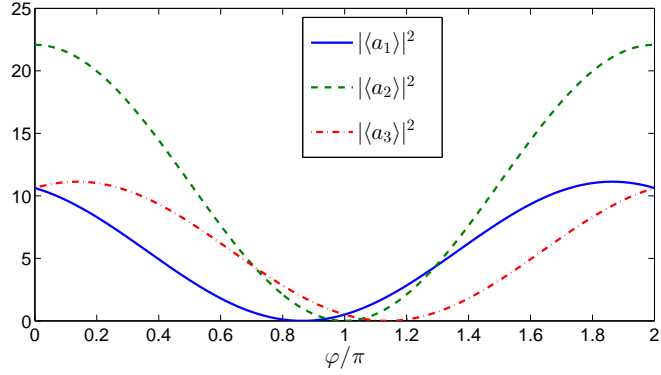}
\caption{The steady state values of the system as a function of the phase difference $\varphi$ between two driving fields in the non-resonant driving situation. The parameters are set as $J=2\gamma,\lambda=2\gamma$, $\Delta=2.773\gamma,\gamma'=0.2\gamma$. Under these parameters, the condition in Eq.~(\ref{condition}) is satisfied.}
\label{steadystate}
\end{figure}

Therefore, by appropriately choosing the parameters so that the condition in Eq.~(\ref{condition})
is satisfied, we can make any of the cavity dark by tuning the phase difference $\varphi$. Under the condition in Eq.~(\ref{condition}), the photonic intensity in each cavity can be expressed in a compact form as
\begin{subequations}
\begin{eqnarray}
|\langle a_1\rangle|^2&=&\frac{\delta^2}{(\Delta^{2}+\gamma^{2}/4)}[1+\cos(\varphi+\theta)],\\
|\langle a_2\rangle|^2&=&\frac{\delta^2}{J^2}(1+\cos\varphi),\\
|\langle a_3\rangle|^2&=&\frac{\delta^2}{(\Delta^{2}+\gamma^{2}/4)}[1+\cos(\varphi-\theta)],
\end{eqnarray}
\label{n123}
\end{subequations}
where $\delta:=\sqrt{2\lambda^{2}J^{2}/(4J^{2}-2\Delta^{2}+\gamma\gamma'/2)}$ and
\begin{equation}
\cos\theta=-\frac{4J^{2}+\gamma\gamma'-4\Delta^{2}}{4J^{2}},\,
\sin\theta=\frac{\Delta(\gamma+\gamma')}{2J^{2}}.
\end{equation}
The phase dependence of steady state values in the case of $\Delta=2.773\gamma$ is shown in Fig.~\ref{steadystate}, where we plot $|\langle a_i\rangle|^{2} (i=1,2,3)$ as a function of $\varphi$. It shows that, $|\langle a_2\rangle|^{2}=0$ when $\varphi=\pi$, and $|\langle a_{1(3)}\rangle|^{2}=0$ when $\varphi\approx0.8(1.2)\pi$. In other words, we can realize the partially dark cavities in the trimer system, which is similar to that in optical molecule or dimer system~\cite{wang2}.
\begin{figure}[tbp]
\centering
\includegraphics[width=0.8\columnwidth]{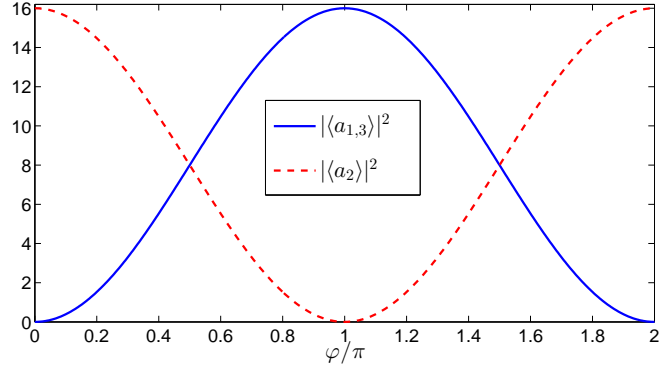}
\caption{The steady state values of the system as a function of the phase difference $\varphi$ between two driving fields in the resonant driving situation. The parameters are set as $J=0.5\gamma,\lambda=2\gamma$, $\Delta=0,\gamma'=0$. Under these parameters, the condition in Eq.~(\ref{condition}) is satisfied.}
\label{steadystateb}
\end{figure}
However, as shown in Fig.~\ref{steadystate}, when any of the cavities is dark with zero average photons in the steady state, the photonic intensity in the other two cavities can not achieve
their maximum values. That is, we can not construct a perfect photonic flash. This defect can be overcome by driving the system resonantly, that is $\Delta=0$, in which situation we need a lossless condition $\gamma'=0$ (whose realization will be discussed later) to grantee the validity of Eq.~(\ref{condition}). As a result, we will reach $\theta=\pi$ and the average photon number become $|\langle a_1\rangle|^{2}=|\langle a_3\rangle|^{2}=2\lambda^2(1-\cos\varphi)/\gamma^2$ and $|\langle a_2\rangle|^{2}=\lambda^2(1+\cos\varphi)/(2J^2)$. In Fig.~\ref{steadystateb}, we plot the curve of the steady values for $\Delta=0$. Therefore, when $\varphi=\pi$, we will have $|\langle a_2\rangle|^{2}=0$, and $|\langle a_1\rangle|^{2}=|\langle a_3\rangle|^{2}$, achieving their maximum values. On contrary, by adjusting $\varphi=0$ (or $2\pi$), the two side cavities in the trimer become completely dark, that is $|\langle a_1\rangle|^{2}=|\langle a_3\rangle|^{2}=0$, and the central cavity becomes bright. In Fig.~\ref{steadystateb}, we have chosen the inter-coupling strength as $J=0.5\gamma$, so that we will reach that the maximum values of $|\langle a_1\rangle|^{2}$, $|\langle a_2\rangle|^{2}$ and $|\langle a_3\rangle|^{2}$ are equal to each other.  In such a way, we have constructed a perfect photonic flash only by tunning the phase difference in an optical trimer system.

\section{Generalization to coupled cavity array}
\label{generalize}
\begin{figure}[tbp]
\centering
\includegraphics[width=0.8\columnwidth]{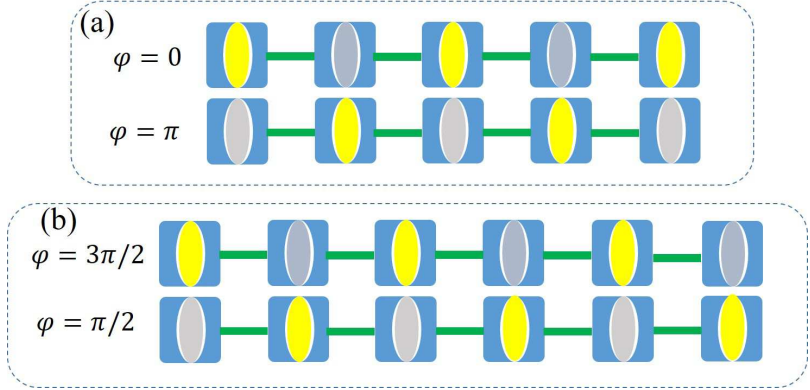}
\caption{The scheme of phase dependent photonic flash based on coupled cavity array system. We take (a) $N=5$  and (b) $N=6$ as the examples for odd and even cavities, respectively.  Here, we use the gray and yellow colors to represent dark and bright cavities respectively.}
\label{flash}
\end{figure}

The scheme of phase dependent photonic flash can also be naturally generalized to the system of $N(>3)$ linearly coupled cavities. By driving the $1$st and $N$th cavities classically, the Hamiltonian of the system reads (in the rotating frame)
\begin{equation}
H=\sum_{i=1}^{N}\Delta_ia_i^{\dagger}a_i+J\sum_{i=1}^{N-1}(a_i^{\dagger}a_{i+1}+h.c.)
+\lambda(a_1e^{i\varphi}+a_N+h.c.).
\end{equation}
Motivated by the realization of perfect photonic flash in optical trimer structure, we here assume a resonant driving situation, and that all of the middle cavities are lossless by setting the parameters as $\Delta_i=0$ and $\gamma_i=\gamma(\delta_{i,1}+\delta_{i,N}),\,(\gamma>0,i=1,2,\cdots,N)$. The Heisenberg-Langevin equation (neglecting the quantum noise) then reads
\begin{subequations}
\begin{eqnarray}
-\frac{\gamma}{2}a_{1}-iJa_{2}&=&i\lambda e^{-i\varphi},\\
a_{i}+a_{i+2}&=&0,\left(i=1,2,...N-2\right),\\
-\frac{\gamma}{2}a_{N}-iJa_{N-1}&=&i\lambda.
\end{eqnarray}
\label{g1}
\end{subequations}
Firstly, let us consider the situation for odd $N$ by assuming $N=2m+1$, where $m$ is an integer. The steady state values can be obtained from Eqs.~(\ref{g1}) as
\begin{eqnarray}
|\langle a_{2j+1}\rangle|^2&=&|\langle a_1\rangle|^2=\frac{2\lambda^2}{\gamma^2}[1+(-1)^{m}\cos\varphi], \\
|\langle a_{2j}\rangle|^2&=&|\langle a_2\rangle|^2=\frac{\lambda^2}{2J^2}[1-(-1)^{m}\cos\varphi],
\end{eqnarray}
for $j=1,2,\cdots m$.  It implies that, when $\cos\varphi=-(-1)^{m}$, all of  the odd number cavities will be completely dark with zero intensity. Meanwhile, the photonic intensity in the even number cavities will achieve their maximums. The opposite event will occur as long as the phase difference satisfies $\cos\varphi=(-1)^m$, for which the even number cavities become dark and the odd ones become bright. In Fig.~\ref{flash}(a), we illustrate the realization of photonic flash in an array of $5\,(m=2)$ coupled  cavities. It shows that the even number cavities will be dark simultaneously (shown by the grey cavities) by tunning the phase difference such that $\varphi=0$, and they will turn bright (shown by the yellow cavities) when $\varphi=\pi$. At the same time, the odd number cavities will experience an exchange from bright ones to dark ones as we adjust the phase difference from $\varphi=0$ to $\varphi=\pi$.

Secondly, let us move to the situation of even cavities, that is $N=2m$, where $m$ is an integer.
A simple calculation shows that the perfect photonic flash can be realized under the parameter
$\gamma=2J$ and the steady state values yield
\begin{eqnarray}
|\langle a_{2j-1}\rangle|^2&=&|\langle a_{1}\rangle|^2=\frac{\lambda^2[1+(-1)^m\sin\varphi]}{2J^2},\\
|\langle a_{2j}\rangle|^2&=&|\langle a_{2}\rangle|^2=\frac{\lambda^2[1-(-1)^m\sin\varphi]}{2J^2},
\end{eqnarray}
where $j=1,2,\cdots,m$. Taking $N=6\, (m=3)$ as an example, we illustrate the realization of
perfect photonic flash in Fig.~\ref{flash}(b). As shown in the figure, the odd (even) number cavities and even (odd) number cavities become bright (dark) alternatively when $\varphi$ experiences a change from $3\pi/2$ to $\pi/2$.

\section{Remark and Conclusion}
\label{remark}
In this paper, we have investigated how to coherently control the steady state and realize the photonic flash in a coupled cavity array system. {All of our discussions are based on the semi-classical approximation, where the decay of the side cavities are taken into consideration, but the effect of the noise is neglected. Here, the decay can be induced by for example the radiative loss and imperfectness of cavities and the noise mainly comes from the quantum fluctuation of the electromagnetic field in the surrounding environment. To go beyond the semi-classical approximation, the noise contribution to the photonic intensity in an optical trimer structure is obtained as}
\begin{equation}
|\langle a_i\rangle_{\rm noi}|^{2}=|\langle a_i\rangle|^2+\sum_{j=1}^{3}n_{j}^{{\rm th}}\int_{0}^{t}d\tau |D_{ij}(\tau-t)|^2.
\label{noise}
\end{equation}
for $i=1,2,3$ and $D(t)=\exp(Mt)$. Here $|\langle a_i\rangle_{\rm noi}|^{2}$ is the photon number by taking the quantum noise into account, $M$ and $|\langle a_i\rangle|^{2}$ are given in Eq.~(\ref{MM}) and  Eqs.~(\ref{n123}), respectively. $n_j^{\rm th}$ is the thermal photon number in the reservoir contact with the $j$th cavity. Therefore the quantum noise at zero temperature ($n_j^{\rm th}=0$) will make no contribution to the function of the photonic flash. Usually, the optical cavities are characterized by their high eigen-frequencies and the noise can be reasonably regarded as at zero temperature. Therefore, we believe that the semi-classical approximation in our discussion will give a logical prediction.

In conclusion, we have theoretically proposed a scheme for photonic flash based on linearly coupled cavity system by only tuning the phase difference between the two driving fields. We firstly demonstrate how to realize a partially dark optical trimer and then design a photonic flash, in which the side cavities are driven resonantly. In such a scheme, the side cavities achieve their vacuum steady state when the central one reaches a maximum photonic intensity, and vice versa. We also generalize the proposal of photonic flash to the system which is composed by multiple coupled cavities, and show that the flash occurs between the odd and even number cavities. In our scheme, the middle cavities are required to be lossless, which can not be achieved strictly. However, within the current technologies, the decay rate can be negligibly small and satisfies our condition approximately.

{At last, we claim that our proposal in this paper maybe of potential application in the quantum device designing and quantum simulation. On one hand, in our current scheme, all of the cavities carry a same frequency, and a natural generalization is to design quantum neon. That is, the resonant frequencies of the cavities are different and so we can control the intensity of light with different ``color'' by adjusting the phase. On the other hand, since the cavities in our system can be dark or bright by tunning the phase difference, we here actually realize a controllable photonic localization phenomenon. Traditionally, the localization is observed in the disordered nonlinear system, such as Anderson localization~\cite{anderson} and many-body localization~\cite{pal}. We hope that our scheme in linear ordered cavity array system will be of potential application.}

\section*{Acknowledgments}
We thank D. L. Zhou and Y. X. Liu for their helpful discussions.

\section*{Funding}
National Natural Science Foundation of China (NSFC) (11404021, 11534002 and 11674049); Jilin
Province Science and Technology Development Plan (20170520132JH); Fundamental Research
Funds for the Central Universities (2412016KJ015 and 2412016KJ004).

\end{document}